\documentclass[review,12pt,authoryear]{elsarticle}

\usepackage{amssymb}
\usepackage{amsmath}
\usepackage{epstopdf}
\usepackage{lineno}
\usepackage{color}
\usepackage{natbib}
\newcommand{\corr}[1]{{\color{black} #1}}

\journal{Geomorphology}

\begin{document}

\begin{frontmatter}



\title{On the effect of two-direction seasonal flows on barchans and the origin of occluded dunes 
\tnoteref{label_note_copyright} \tnoteref{label_note_doi}
} 

\tnotetext[label_note_copyright]{\copyright 2024. This manuscript version is made available under the CC-BY-NC-ND 4.0 license http://creativecommons.org/licenses/by-nc-nd/4.0/}

\tnotetext[label_note_doi]{Accepted Manuscript for Geomorphology, accepted, 2024, DOI: 10.1016/j.geomorph.2024.109488}


\author[aff1]{Willian R. Assis}
\author[aff2]{Danilo S. Borges}
\author[aff2]{Erick M. Franklin\corref{cor1}}
\ead{erick.franklin@unicamp.br}
\cortext[cor1]{Corresponding author. phone: +55 19 35213375}

\affiliation[aff1]{organization={Saint Anthony Falls Laboratory, University of Minnesota},
	addressline={2 3rd Ave SE}, 
	city={Minneapolis},
	postcode={55414}, 
	state={Minnesota},
	country={USA}}

\affiliation[aff2]{organization={School of Mechanical Engineering, University of Campinas - UNICAMP},
	addressline={Rua Mendeleyev, 200}, 
	city={Campinas},
	postcode={13083-860}, 
	state={SP},
	country={Brazil}}

\begin{abstract}
We inquire into the morphodynamics of barchans under seasonal flows. For that, we carried out grain-scale numerical computations of a subaqueous barchan exposed to two-directional flows, and we varied the angle and frequency of oscillations. We show that when the frequency is lower than the inverse of the characteristic time for barchan formation, the dune adapts to the new flow direction and recovers the barchan shape while losing less grains than under one-directional flow. For higher frequencies, the dune has not enough time for adaptation and becomes more round while losing more grains. For both cases, we show, for the first time, the typical dynamics of grains (trajectories and forces). In particular, the round barchans are similar to the so-called occluded dunes observed on Mars, where seasons have very high frequencies compared to the dune timescale, different from Earth. Our results represent a possible explanation for that shape.
\end{abstract}

\end{frontmatter}




\section{Introduction}

Crescent-shaped dunes, known as barchans, are ubiquitous on Earth, Mars and other celestial bodies  \citep{Bagnold_1, Herrmann_Sauermann, Hersen_3, Elbelrhiti, Claudin_Andreotti, Parteli2, Courrech}, where, although growing under generally one-directional flows, are exposed to seasonal winds \citep[\corr{variations in direction and strength as seasons change,}][]{Bagnold_1, Pye, Gao2}. Given their alignment in the main flow direction, and the long timescales of eolian and Martian barchans \citep[of the order of years and up to milleniums, respectively,][]{Claudin_Andreotti} the vast majority of studies investigated barchans under one-directional flows. However, \citet{Parteli5} and \citet{Taniguchi} showed that bimodal winds can change drastically the shape of barchans, even giving origin to other types of dunes.

One of the first experiments investigating how two-directional winds change ripple patterns was carried out by \citet{Rubin}. For that, they used a board covered with sand which they exposed to wind and rotated as desired, and found that the angle $\alpha$ between wind directions (divergence angle) and the transport ratio $R$ between the flows in each direction result in transverse, oblique, and longitudinal ripples. In particular, they found that symmetric flows tend to produce transverse ripples when $\alpha$ $<$ 90$^{\circ}$ and longitudinal ripples when $\alpha$ $>$ 90$^{\circ}$ (for $\alpha$ varying within 0 and 180$^{\circ}$). Although not specifically for dunes, their results shed light on the role of the divergence angle on bedform orientation. Later, \citet{Parteli5} investigated the morphology of dunes under bimodal winds by adapting a continuum saltation model \citep{Sauermann_4, Kroy_C, Kroy_A}. They observed that when the divergence angle $\alpha$ is greater than 90$^\circ$, dunes tend to be aligned in the longitudinal direction \citep[in agreement with the results for ripples obtained by][]{Rubin}, independent of the sand availability and atmosphere. They also showed that increasing sand availability leads to seif dunes, while variations in the angle and frequency of oscillation when the quantity of sand is limited lead to unusual shapes, such as the \corr{occluded barchans (round dunes with a slip face and horns)} when 40$^{\circ}$ $<$ $\alpha$ $<$ 80$^{\circ}$ and drop barchans when $\alpha$ $\approx$ 120$^{\circ}$, both similar to bedforms observed on Mars. \corr{In addition, they found that dome dunes (no slip face), also observed on Earth, appear for $\alpha$ $<$ 90$^{\circ}$ and high frequencies (ratio between the period of oscillations and dune migration time of the order of 10$^{-4}$).} The explanation advanced by the authors for the appearance of those forms concerns local erosion and deposition, so that measurements at the scale of grains are crucial for thoroughly understanding the problem (as recognized by the authors). Later, \citet{Taniguchi} carried out experiments in a water flume for investigating the arising patterns when an isolated dune was under bimodal flows. They identified different patterns, showed that they depend basically on the divergence angle $\alpha$ and flow strength (or duration) of each mode, and proposed a map for classifying patterns on the angle-strength space. The proposed map is a significant advance for comprehending the conditions for pattern formation, and agrees well with field data from Western Sahara (although the map does not consider sand availability); however, detailed measurements of morphology and motion of grains were not conducted.

Based on subaqueous experiments, \citet{Courrech2} proposed a model for predicting the orientation of dunes under multidirectional winds, and showed that different orientations can appear under bimodal winds. They showed, however, that dune orientation depends strongly on sand availability: in the absence of sand limitation, dunes are oblique to the direction of sand transport and form barchanoid ridges, while when sand is limited longitudinal or seif dunes appear, in contradiction, in part, with the results of \citet{Parteli5}. \citet{Gadal} carried out a linear stability analysis (LSA) of sand beds under two-directional winds, and found that a transition from transverse to longitudinal dunes occurs, depending on the transport ratio $R$ and angle $\alpha$ between flow directions. For bimodal winds of \corr{the} same strength, their results agree with those of \citet{Rubin} and \citet{Parteli5}: the transition from transverse to longitudinal dunes occurs for $\alpha$ = 90$^\circ$, and, in addition, they showed that the wavelength $\lambda$ of transverse dunes increases and that of longitudinal dunes decreases with increasing $\alpha$. The results for transverse dunes were validated with experiments carried out under water, but not those for longitudinal dunes. As remarked by the authors, discrepancies between the experiments and LSA can be caused by nonlinearities. Numerical simulations at mesoscale (smaller than the bedform but much larger than the diameter of grains) were also carried out under two- \citep{Gao} and three-direction \citep{Rubin2} flows and different availability of sand by using a cellular automaton
model \citep{Narteau}. For two-direction flows, \citet{Gao} showed that for large sand availability (erodible bed) only linear dunes grow, with the wind direction and strength controlling dune orientation, while for small availability (non-erodible ground) there is a transition from linear to barchan dunes as the amount of available sand decreases, which depends also of the wind regime. For three-direction winds, \citet{Rubin2} showed that sand availability is the main parameter controlling the dune morphology, with long-crest periodic patterns growing over erodible grounds and different shapes (including barchans) over non-erodible grounds, in agreement with the findings of \citet{Gao}. \corr{\citet{Gao3} carried out the same type of simulation, but varying the standard deviation of sand flux orientation (instead of wind direction), and showed that larger deviations can account for the transition from barchan to dome dunes \citep[round dunes without slip face and horns,][]{Courrech3}.}  More recently, \citet{Gao2} investigated the extreme case of dunes under reversing flows ($\alpha$ $=$ 180$^\circ$) found at the border between the Tibetan Plateau and the Taklamakan Desert, in particular the migration of dunes in the direction opposite to that of the resultant sand transport. The authors showed that a highly non-linear behavior arises from transients during flow reversals, increasing the transport rates on the crest region while it migrates toward the new downstream direction with a celerity that depends on the flow speed up, explaining, thus, the apparent reverse migration.

The previous results shed light on how the dune morphology changes with variations in the flow direction; however, they were obtained at the bed scale only, and information at the grain scale is still missing, such as the number of grains lost by dunes and grains' trajectories. Besides, for more complex dunes, such as barchans, the existence of strong nonlinearities may invalidate the use of LSA. In the particular case of barchan dunes, the scales observed under water are much smaller and faster in comparison with their eolian and Martian counterparts \citep[centimeters and minutes under water, \corr{and up to one kilometer and millenniums on Mars,}][]{Claudin_Andreotti}, while their shape remains roughly the same \citep{Hersen_1}. Taking advantage of such faster scales, previous works carried out experiments in subaqueous environment to investigate the growth and evolution of barchans \citep{Hersen_1, Alvarez, Alvarez3, Alvarez4}, as well as barchan-barchan interactions \citep{Assis, Assis2, Assis3}, but very few investigated changes in the flow direction \citep[for example,][]{Hersen_6, Taniguchi}. The same applies to numerical simulations at the grain scale, where, to the best of our knowledge, the only grain scale simulations of barchans are those performed by \citet{Alvarez5, Alvarez7} and \corr{\citet{Lima2, Lima3}}, all of them for one-direction flows. In those simulations, the authors made use of DEM (discrete element method) for the grains and LES (large eddy simulation) for the fluid, where the smaller scales resolved for the fluid were of the order of the grain diameter. Therefore, information impossible to be obtained from experiments were computed numerically, such as the motion and resultant force for each individual grain. The same technique can be applied, thus, to investigate barchans under bimodal flows.

In this paper, we inquire into the morphodynamics of subaqueous barchans under bimodal flows. For that, we carried out numerical computations in which a barchan dune is exposed to changes in the flow direction, and we varied both the angle (limited to $\alpha$ $\leq$ 90$^\circ$ and $\alpha$ $=$ 180$^\circ$) and frequency of oscillations. \corr{We could perform computations at the grain scale because subaqueous barchans are much smaller and faster (consisting of around 10$^5$ particles and taking minutes to develop) than their terrestrial (eolian) and Martian counterparts (quadrillions of particles for a terrestrial desert dune, and even more for Martian dunes), while keeping roughly the same morphology \citep{Hersen_1, Claudin_Andreotti}.} We used CFD-DEM (computational fluid dynamics - discrete element method), where the flow is solved using LES and the motion of each grain is computed at each time step. Different from previous works on bimodal winds, we solve the problem at the grain scale, and investigate, in especial, the morphology of barchans and the motion of their grains. We observe that when the frequency of oscillation is lower than the inverse of the characteristic time for barchan formation, the dune adapts to the new flow direction and recovers the barchan shape while losing less grains, for all angles tested. For higher frequencies, the dune has not enough time to adapt and gives origin to more round barchans that lose more grains. Finally, we show the trajectories of grains, the resultant forces on grains, the number of grains that the barchan loses, and how its size decreases along time. \corr{Some of these data are unfeasible to be obtained from experiments or field measurements, especially the resultant force acting on each grain.} In particular, \corr{we note that} the round barchans are similar to the occluded dunes observed on Mars, where seasons have very high frequencies compared to the dune timescale, different from Earth. Our results represent, thus, a possible explanation for that shape.

\section{Materials and Methods}

\begin{sloppypar}
	We carried out CFD-DEM computations, in which we made use of LES for solving the fluid flow. For that, we used the open-source code \mbox{CFDEM} \citep[www.cfdem.com,][]{Goniva}, coupling the open-source CFD code OpenFOAM with the open-source DEM code LIGGGHTS \citep{Kloss, Berger}. In general terms, the CFD part computes the mass (Eq. \ref{mass}) and momentum (Eq. \ref{mom}) equations for the fluid,
\end{sloppypar}

\begin{equation}
	\nabla\cdot\vec{u}_{f}=0 \, ,
	\label{mass}
\end{equation}

\begin{equation}
	\frac{\partial{\rho_{f}\vec{u}_{f}}}{\partial{t}} + \nabla \cdot (\rho_{f}\vec{u}_{f}\vec{u}_{f}) = -\nabla P + \nabla\cdot \vec{\vec{\tau}} + \rho_{f}\vec{g} - \frac{N}{V}\vec{F}_{fp} \, ,
	\label{mom}
\end{equation}

\noindent where $\vec{g}$ is the acceleration of gravity, $\vec{u}_{f}$ is the fluid velocity, $\rho_{f}$ is the fluid density, $P$ the fluid pressure, $\vec{\vec{\tau}}$ the deviatoric stress tensor of the fluid, $\vec{F}_{fp}$ is the resultant of fluid forces acting on each grain, $N$ is the number of grains in a given cell and $V$ is the cell volume. In $\vec{F}_{fp}$, we considered the fluid drag, the stress tensor, and the virtual mass, more details are available in the Supplementary Material and \citep{Lima2}. The DEM part solves the linear (Eq. \ref{Fp}) and angular (Eq. \ref{Tp}) momentum equations for each solid particle,

\begin{equation}
	m_{p}\frac{d\vec{u}_{p}}{dt}= \vec{F}_{p}\, ,
	\label{Fp}
\end{equation}

\begin{equation}
	I_{p}\frac{d\vec{\omega}_{p}}{dt}=\vec{T}_{c}\, ,
	\label{Tp}
\end{equation}

\noindent where, for each grain, $m_{p}$ is the mass, $\vec{u}_{p}$ is the velocity, $I_{p}$ is the moment of inertia, $\vec{\omega}_{p}$ is the angular velocity, $\vec{T}_{c}$ is the resultant of contact torques between solids \citep[we neglect torques caused directly by the fluid in the angular momentum, since those due to contacts are much higher,][]{Tsuji, Tsuji2, Liu}, and $\vec{F}_{p}$ is the resultant force (weight, contact and fluid forces), given by

\begin{equation}
	\vec{F}_{p}= \vec{F}_{fp} + \vec{F}_{c} + m_{p}\vec{g}\, ,
	\label{Fp2}
\end{equation}

\noindent where $\vec{F}_{c}$ is the resultant of contact forces between solids \citep[more details for computing $\vec{F}_{c}$  are available in the Supplementary Material and][]{Lima2}.

In our simulations, after having formed a developed barchan from an initially conical heap \citep[as in][]{Lima2}, we imposed changes in the flow direction by rotating instantaneously the dune (and deleting the grains that had left the dune and been entrained further downstream). The changes were either constant (fixed) or oscillated with a given frequency and angle, which we varied in the different runs. The CFD domain is a 3D channel of size $L_x$ = 0.4 m, $L_y$ = $\delta$ = 0.025 m and $L_z$ = 0.1 m, where $x$, $y$ and $z$ are the longitudinal, vertical and spanwise directions, respectively. The channel Reynolds number based on the cross-sectional mean velocity $U$, Re = $U 2\delta \nu^{-1}$, is 14,000, and the Reynolds number based on shear velocity $u_*$, Re$_*$ = $u_* \delta \nu^{-1}$, is 400, where $\nu$ is the kinematic viscosity of the fluid (water in our simulations). The granular material forming the initial heap consisted of 10$^5$ glass spheres (density $\rho_p$ = 2500 kg/m$^3$), whose sizes were randomly distributed within 0.15 mm $\leq$ $d$ $\leq$ 0.25 mm following a Gaussian distribution. \corr{The fluid properties were those of water, which allowed the use of a small system for the grains (otherwise, for air the number of particles would be of the order of one quadrillion, which is unfeasible in terms of CFD-DEM simulations). Although the motion of grains over the dune is different in each case \citep[saltation in the eolian case and rolling/sliding in the subaqueous case,][]{Alvarez3}, the aspect ratio of morphological dimensions remains approximately the same \citep{Hersen_1}. However, extrapolations to dunes on Mars or terrestrial deserts must be done with care.} A description of the fundamental and implemented equations, CFD meshes and convergence, DEM parameters, and tests can be found in \citet{Lima2}, and more details of the numerical setup are available in the Supplementary Material. In addition, files of the numerical setup are available in an open repository \citep{Supplemental2}.

\section{Results}

\subsection{Morphology}

\begin{figure}[ht]
	\begin{center}
		\includegraphics[width=0.7\linewidth]{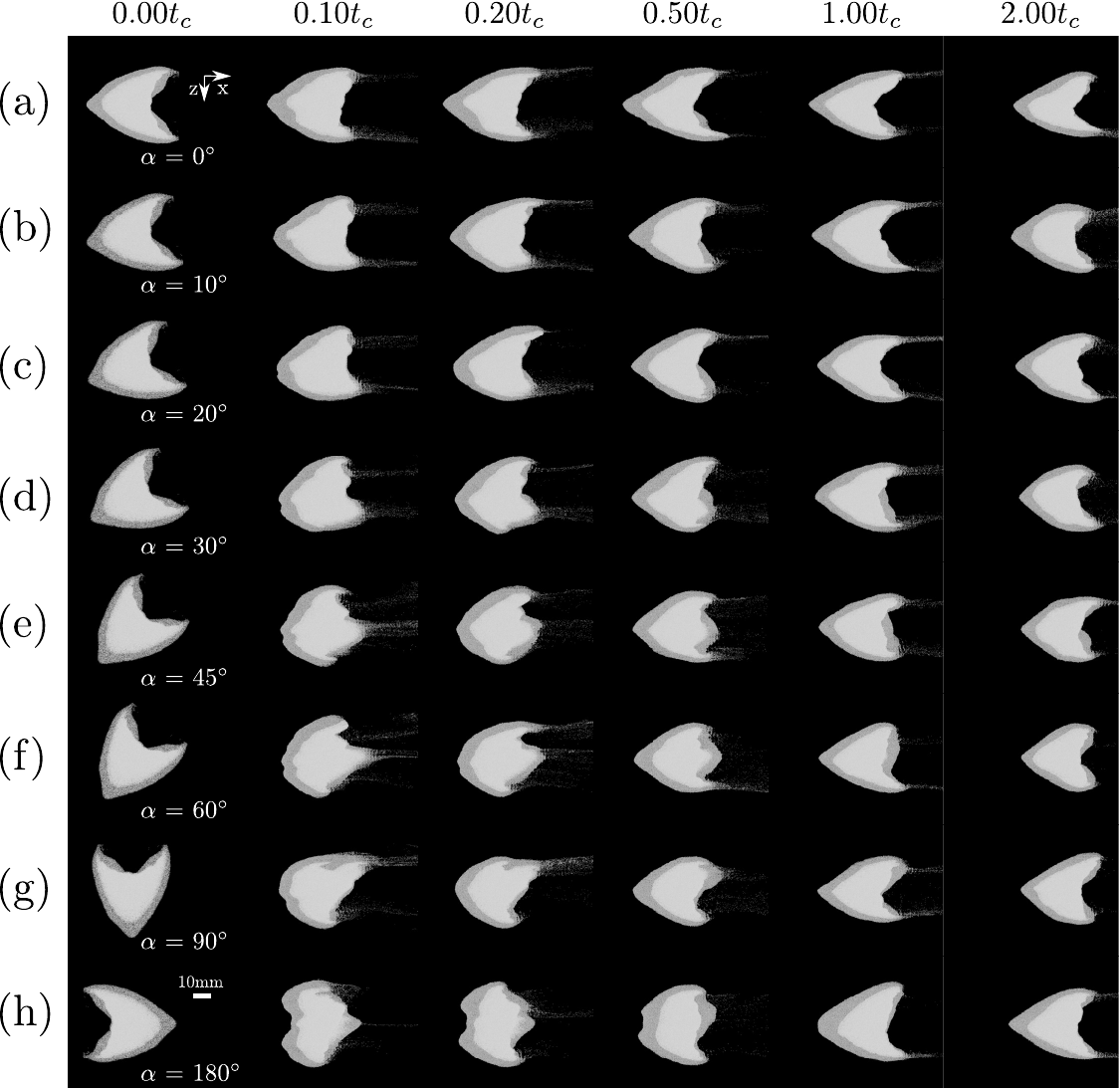}\\
	\end{center}
	\caption{Snapshots of a barchan dune under a flow oscillating with $f$ = 0/$t_c$ (the new flow direction is kept constant) and different divergence angles: (a) $\alpha$ $=$ 0$^\circ$, (b) $\alpha$ $=$ 10$^\circ$, (c) $\alpha$ $=$ 20$^\circ$, (d) $\alpha$ $=$ 30$^\circ$, (e) $\alpha$ $=$ 45$^\circ$, (f) $\alpha$ $=$ 60$^\circ$, (g) $\alpha$ $=$ 90$^\circ$, and (h) $\alpha$ $=$ 180$^\circ$.}
	\label{fig:snapshots1}
\end{figure}

\begin{figure}[ht]
	\begin{center}
		\includegraphics[width=1\linewidth]{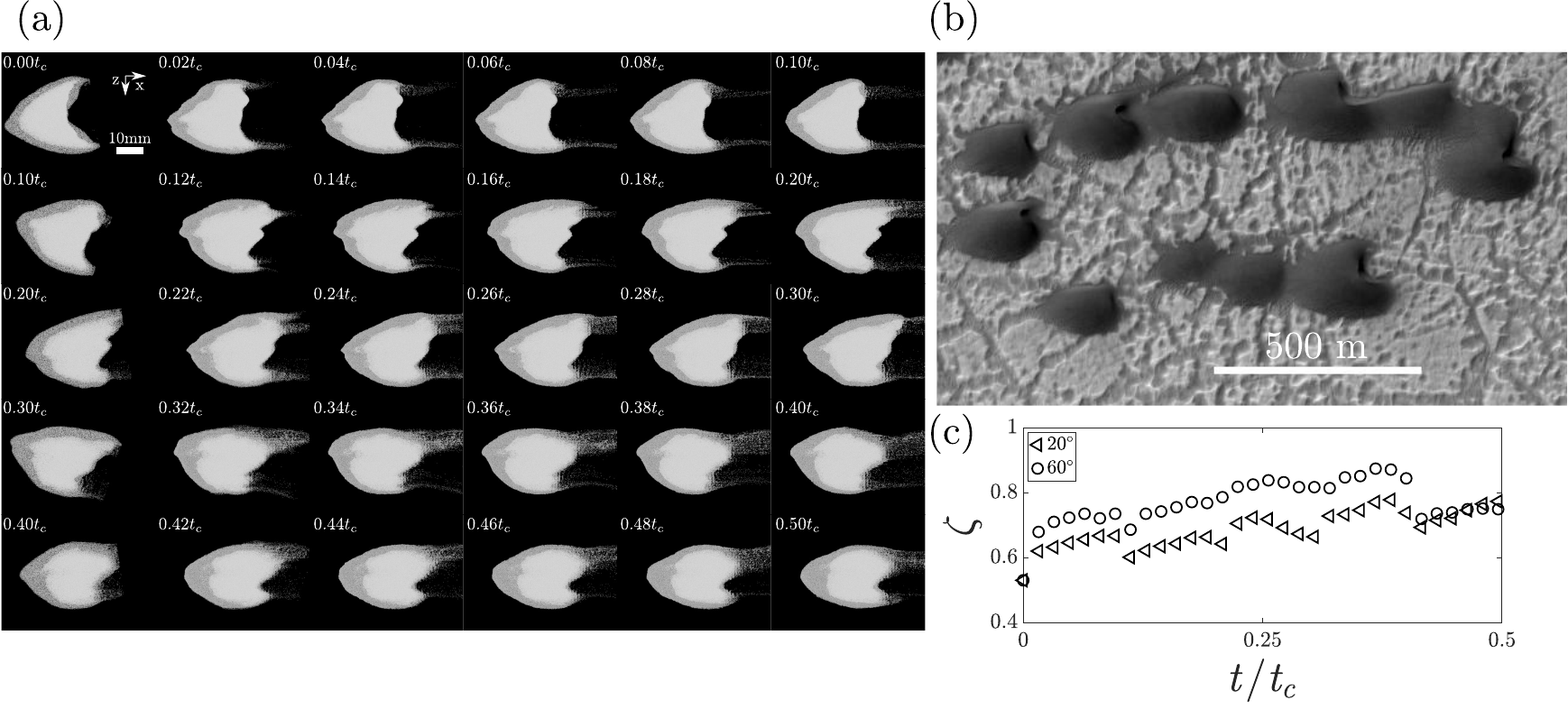}\\
	\end{center}
	\caption{(a) Snapshots of a barchan dune under a two-direction flow oscillating with $f$ $=$ 10/$t_c$ and $\alpha$ $=$ 20$^{\circ}$. (b) Barchan (occluded) dunes on Mars: Latitude (centered) -39.524$^\circ$, longitude (east) 355.402 $^\circ$, September 2013 (Courtesy: NASA/JPL-Caltech/UArizona). (c) Time evolution of the bedform roundness $\zeta$ for $f$ $=$ 10/$t_c$ (values of $\alpha$ are shown in the key).}
	\label{fig:snapshots2}
\end{figure}

We observe two distinct behaviors: (i) when the new flow direction remains constant or the period of flow oscillation is higher than the characteristic time for the growth of a barchan dune, $1/f$ $>$ $t_c$, the barchan adapts to the new flow condition and recovers its shape, which becomes aligned in the new flow direction \citep[see the Supplementary Material, or][for a description of the characteristic time $t_c$, which is 49.8 s in the present case]{Alvarez}; (ii) when the oscillation period is lower than the barchan adaptation time, $1/f$ $<$ $t_c$, the barchan has not enough time to adapt to the new flow direction and, along the cycles, becomes more round. For example, Figs. \ref{fig:snapshots1} and \ref{fig:snapshots2} show snapshots of a barchan dune under new flow directions for different angles and frequencies.

Figure \ref{fig:snapshots1} shows snapshots of a barchan under a flow that remains constant after the flow direction has changed (computed until 2$t_c$), corresponding thus to the $1/f$ $>$ $t_c$ case. In the figure, panels (a) to (h) correspond to angles increasing from $\alpha$ $=$ 0$^\circ$ to $\alpha$ $=$ 180$^\circ$, and we observe that, as expected, the barchans have recovered their shape at $t$ = $t_c$ for all angles. During the transient ($t$ $<$ $t_c$), the grains move over the dune to form the new lee face and horns (as can be seen in the movies available in the Supplementary Material). As a consequence, the ratio of the width $W$ to length $L$ of barchans oscillates around the values found for a barchan that hasn't undergone changes in direction, as shown in Fig. \ref{fig:grain_loss}e, which presents the time evolution of $W/L$ for the different deviation angles simulated. We notice that within $t$ $<$ $t_c$, barchans under large deviation angles ($\alpha$ $\geq$ 60$^{\circ}$) have ratios $W/L$ that diverge more expressively from the $\alpha$ $=$ 0$^\circ$ case (deviations reaching 60\%), and that from $t$ $=$ $t_c$ on all cases tend to converge to the same value, eventually reaching 0.8 $\leq$ $W/L$ $\leq$ 0.9 ($W/L$ $\approx$ 0.8 for the case $\alpha$ $=$ 0$^\circ$ case at all times). The formation of the new lee face and horns during the transient ($t$ $<$ $t_c$) could affect the dune asymmetry, \corr{especially} for angles approaching 90$^\circ$. The asymmetry can be expressed in terms of the ratio between the left and right horns, $L_{hl}$ and $L_{hr}$, respectively, as shown in Fig. \ref{fig:grain_loss}g as a function of time. From this figure, we observe that $L_{hl}/L_{hr}$ deviates considerably from unity for all angles ($\alpha$ = 0$^{\circ}$ included), oscillating with a period that seems to scale with $t_c$. These oscillations are a consequence of the intermittent motion of barchans (successive accumulation and avalanching of grains near the crest), and a behavior similar to the $\alpha$ $=$ 0$^{\circ}$ case shows that barchans have adapted to the new flow condition.

Figure \ref{fig:snapshots2}a shows snapshots of a barchan dune under a bimodal flow oscillating with $f$ $=$ 10/$t_c$ and $\alpha$ $=$ 20$^{\circ}$, corresponding thus to the $1/f$ $<$ $t_c$ case. We observe that during each cycle the barchan does not have enough time to adapt to the new flow condition and, as the total time evolves, becomes more round, eventually reaching an occluded shape \citep{Parteli5}, similar to barchans found on the surface of Mars (some examples are shown in Fig. \ref{fig:snapshots2}b). This is reflected in the roundness $\zeta$ of bedforms, computed as

\begin{equation}
	\zeta = \frac{4\pi A}{s^2} \left(1 - \frac{0.5}{\beta} \right)^2 \,\,,
	\label{eq_roudness}
\end{equation}

\noindent and presented in Fig. \ref{fig:snapshots2}c. In Eq. \ref{eq_roudness}, $A$ is the surface area projected in the horizontal direction, $s$ the corresponding perimeter, and $\beta$ = $s/(2 \pi)$ $+$ 0.5, so that $\zeta$ tends to unity as the bedform projection approaches a perfect circle. From Fig. \ref{fig:snapshots2}c, we observe that the roundness of barchans increases by roughly 30\% as time goes on (from $\zeta$ $\approx$ 0.6 to $\zeta$ $\approx$ 0.8), corroborating the fact the barchans become more round when $1/f$ $<$ $t_c$ (in the $f$ $=$ 0/$t_c$ case, 0.5 $\lesssim$ $\zeta$ $\lesssim$ 0.7 when $t$ $>$ $t_c$, see the Supplementary Material). In terms of asymmetry, Fig. \ref{fig:grain_loss}h shows that, along cycles, $L_{hl}/L_{hr}$ reaches higher and lower values (within approximately 0.7 and 1.1), and tends to a value closer to unity (within 0.8 and 0.9) as cycles go on. This corroborates that the barchan has not enough time to adapt to the new flow direction before the new change takes place, and assumes a more round shape.

\subsection{Mass variation and grain trajectories}
\label{subsection_grains}

\begin{figure}[!h]
	\begin{center}
		\includegraphics[width=0.9\linewidth]{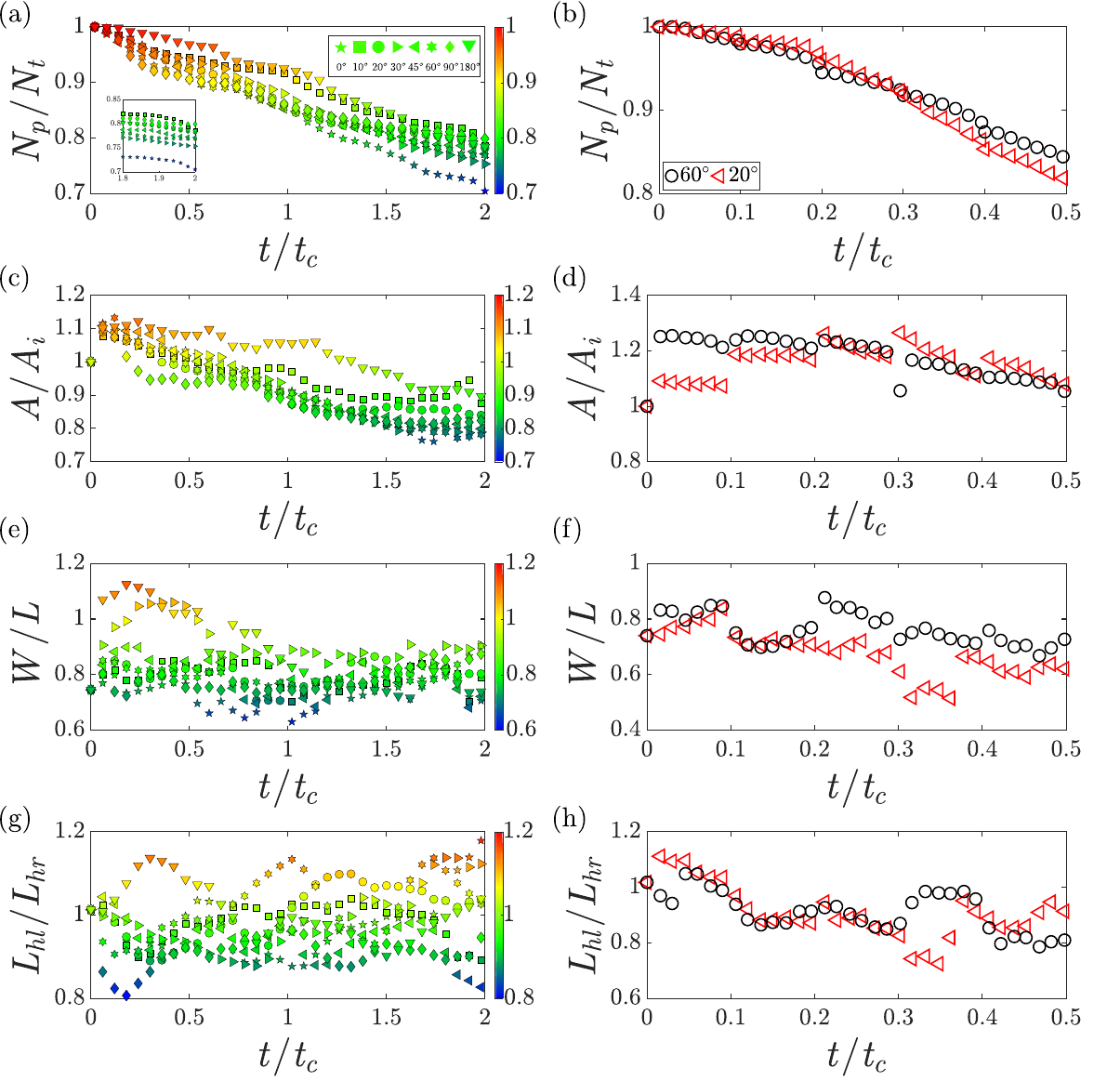}\\
	\end{center}
	\caption{Size reduction and morphology variations along time for different frequencies and angles. (a) and (b) Number of particles in a dune $N_p$ normalized by the initial number of particles $N_t$, for $f$ = 0/$t_c$ and $f$ = 10/$t_c$, respectively. (c) and (d) Surface area of dunes projected in the horizontal direction $A$ normalized by the initial area $A_i$, for $f$ = 0 and $f$ = 10/$t_c$, respectively. (e) and (f) Ratio between the barchan width $W$ and length $L$, for $f$ = 0/$t_c$ and $f$ = 10/$t_c$, respectively. (g) and (f) Length of the left horn divided by that of the right horn, $L_{hl}/L_{hr}$, for $f$ = 0 and $f$ = 10/$t_c$, respectively. In the panels for $f$ = 0/$t_c$, values can be read on the colorbar on the right of each panel, and for $f$ = 10/$t_c$ the angles are represented by symbols shown in the key of panel (a).}
	\label{fig:grain_loss}
\end{figure}

We computed the time evolution of the number of grains that remain in the dune, $N_p$, and the surface area of dunes projected in the horizontal direction, $A$, after the change in the flow direction has taken place. Figures \ref{fig:grain_loss}a and \ref{fig:grain_loss}b show the number of particles in a dune $N_p$ normalized by the initial number of particles $N_t$, for $f$ = 0/$t_c$ and $f$ = 10/$t_c$, respectively. At the first order, we observe that under high-frequency oscillations ($1/f$ $<$ $t_c$) barchans lose more grains than under low-frequency oscillations (or continuous flow): while at $t$ = 0.5$t_c$ barchans under $f$ = 10/$t_c$ keep only 80-85\% of their initial particles, those under $f$ = 0/$t_c$ keep more than 90\% of their initial grains (note that simulations for $f$ = 0/$t_c$ were longer than those for $f$ = 10/$t_c$). At the second order, we notice that barchans undergoing a flow change that remains constant for more than $t_c$ lose less grains than when no change in direction has occurred ($\alpha$ = 0$^\circ$): while in the latter case the barchan keeps 70\% of its grains, in the former they keep between 75 and 83\% of their grains by the end of 2$t_c$ (as can be seen on the inset of Fig. \ref{fig:grain_loss}a). Finally, we observe some oscillations in $N_p$ over time, meaning that grains are lost with some intermittency. For $f$ = 10/$t_c$, the $N_p$ oscillations correspond exactly to the period of flow changes (0.1$t_c$), while for $f$ = 0/$t_c$ the system selects a proper oscillation that scales with $t_c$, i.e., $f$ $\sim$ $1/t_c$. Because data between the different angles are out-of-phase, it is difficult to \corr{ascertain} for which angle the dune keeps more grains.

Figures \ref{fig:grain_loss}c and \ref{fig:grain_loss}d show the projected area $A$ normalized by its initial value $A_i$, for $f$ = 0/$t_c$ and $f$ = 10/$t_c$, respectively. We observe in both cases oscillations that follow closely those for the loss of grains, but the behavior is not the same: $A$ decreases (in average) along time in the continuous case, while it initially increases and then remains roughly constant when $f$ = 10/$t_c$ (until $t$ = 0.5$t_c$, end of our simulations). For the latter, the increase in $A$ is a consequence of the horizontal spreading of grains, making the barchan more round in the horizontal plane and leading it to the occluded shape, as shown in Fig. \ref{fig:snapshots2}a. \corr{We note that the resulting occluded shape has still a slip face and horns, different from the dome dunes appearing also under oscillating flows \citep{Gao3, Courrech3}.} For the continuous case, we observe that as $t$ tends to 2$t_c$ the area tends to be higher for barchans that undergo a change in direction (with respect to the case $\alpha$ = 0$^\circ$), but the oscillations in $A$, however, hinder further conclusions on how it varies with $\alpha$.

\begin{figure}[!h]
	\begin{center}
		\includegraphics[width=0.99\linewidth]{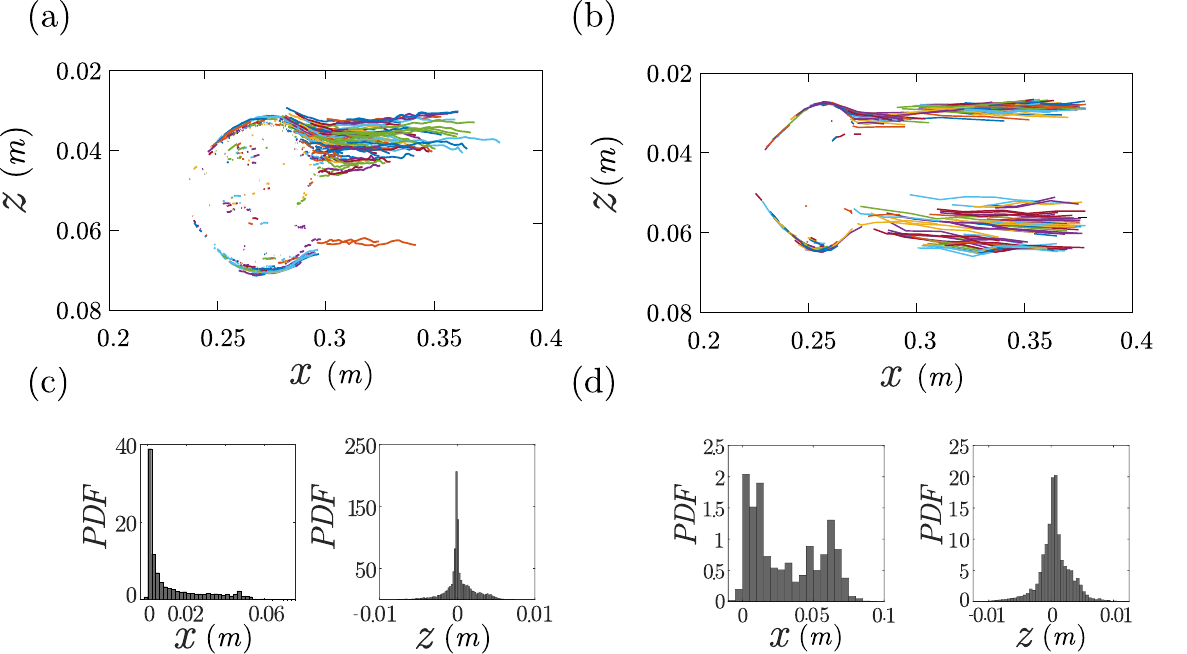}\\
	\end{center}
	\caption{Top view of the trajectories of moving grains for $\alpha$ = 20$^\circ$, and (a) $f$ = 10/$t_c$ and (b) $f$ = 0/$t_c$. Trajectories were plotted from $t$ =  0.4$t_c$ to 0.5$t_c$, and we only tracked grains for which $|\vec{u}_p|$ $>$ 0.1$u_*$. Different colors were used for better discrimination of trajectories. For $f$ = 10/$t_c$ and $f$ = 0/$t_c$, respectively, panels (c) and (d) present the probability density functions (PDFs) of the longitudinal and transverse displacements of tracked particles in the considered interval.}
	\label{fig:trajectories}
\end{figure}

\begin{figure}[!h]
	\begin{center}
		\includegraphics[width=0.6\linewidth]{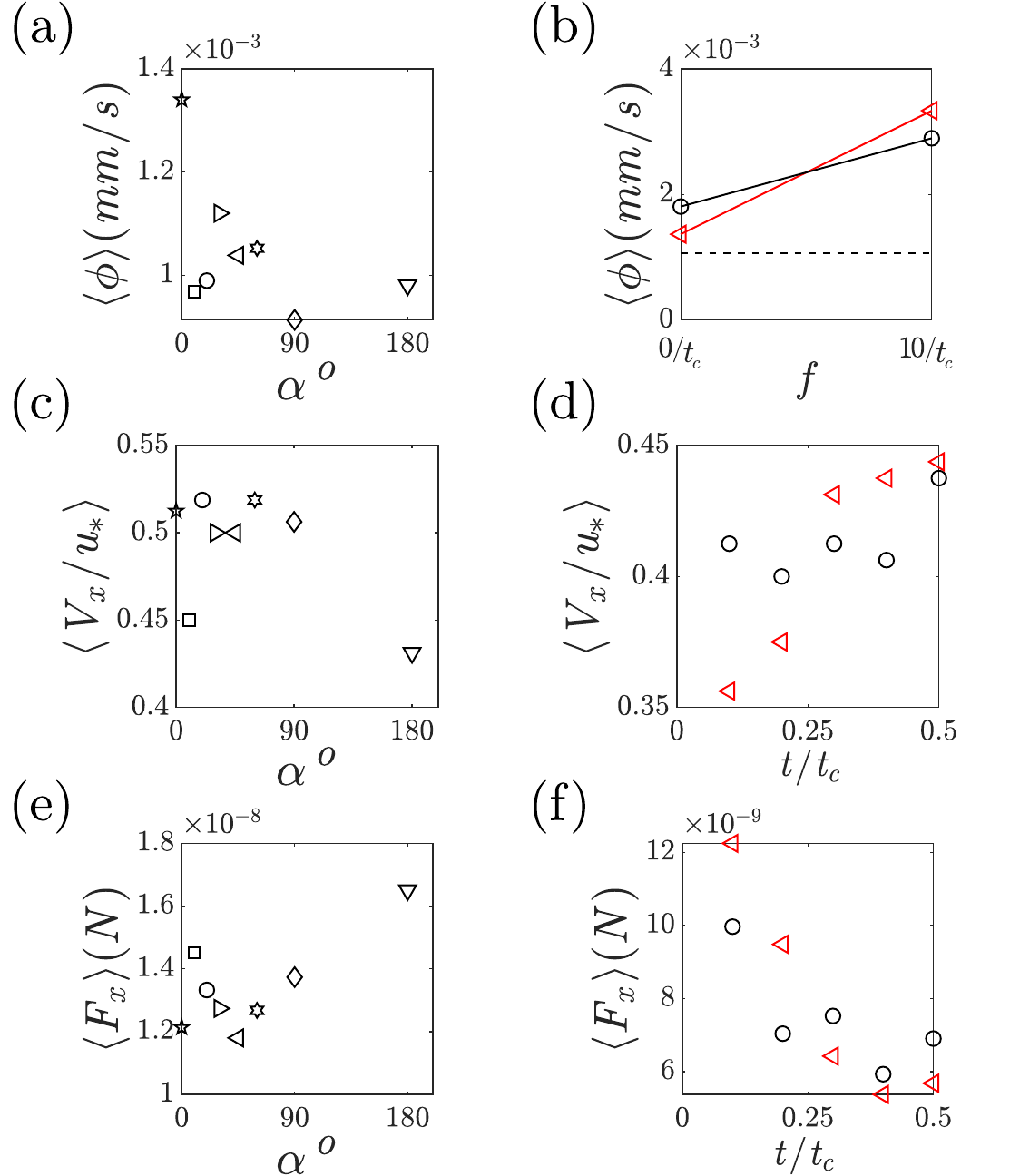}\\
	\end{center}
	\caption{(a) Time-averaged flux of grains $\left< \phi \right>$ lost by the barchan dune as a function of  the divergence angle $\alpha$, for $f$ = 0/$t_c$, (b) $\left< \phi \right>$ over the barchan dune as a function of the frequency $f$, for $f$ = 0/$t_c$ and $f$ = 10/$t_c$. (c) Time-ensemble average of the longitudinal component of the grains' velocity $\left< v_x \right>$ as a function of $\alpha$, for $f$ = 0/$t_c$. (d) $\left< v_x \right>$ (averages computed for each cycle) as a function of time, for $f$ = 10/$t_c$. (e) Time-ensemble average of the longitudinal component of the resultant force $\left< F_x \right>$ as a function of $\alpha$, for $f$ = 0/$t_c$. (f) $\left< F_x \right>$ (averages computed for each cycle) as a function of time , for $f$ = 10/$t_c$.  In panels (b), (d) and (f), the red triangles correspond to $\alpha$ = 20$^\circ$ and the black circles to  $\alpha$ = 60$^\circ$, in panel (b) the dashed line corresponds to $\left< \phi \right>$ under continuous flow ($\alpha$ = 0$^\circ$) and at $t$ = 0.5$t_c$, and in panels (c) and (d) velocities were normalized by $u_*$.}
	\label{fig:grain_motion}
\end{figure}

We computed also the trajectories of grains, the corresponding velocities and fluxes, and the resultant force acting on each grain (see the Supplementary Material for movies showing the motion of grains over and around the dune). In the following, the $x$ subscript corresponds to the longitudinal direction (with reference to the fluid flow), and the $z$ subscript to the transverse direction. Figures \ref{fig:trajectories}a and \ref{fig:trajectories}b show top views of the trajectories of moving grains for $f$ = 10/$t_c$ and $f$ = 0/$t_c$, respectively, and $\alpha$ = 20$^\circ$. For that, we tracked grains for which $|\vec{u}_p|$ $>$ 0.1$u_*$, and considered an interval of 0.1$t_c$ (from $t$ = 0.4$t_c$ to 0.5$t_c$) for plotting the trajectories in order to compare directly one cycle of the oscillating case with the continuous one. Basically, we can observe a greater asymmetry in the $f$ = 10/$t_c$ case, which repeats at each change of direction (each cycle, although not shown in the figure), indicating a stronger change of shape. We computed the probability density functions (PDFs) of typical distances traveled by grains (between starting moving and stopping), which are shown in Figs. \ref{fig:trajectories}c and \ref{fig:trajectories}d. From the PDFs, we observe that the typical distances traveled in the longitudinal and transverse directions are 30.6 and 0.6 mm for the continuous case, and 11.6 and 0.3 mm for the oscillating case, showing, thus, distances 160\% larger in the longitudinal and 100\% in the transverse direction in the continuous case. Therefore, the displacement distances within cycles of the fast-oscillating case are smaller when compared with the continuous case, the grains being under stronger changes of entraining forces in the former case.

Figure \ref{fig:grain_motion}a shows the time-averaged flux $\left< \phi \right>$ of grains lost by the barchan as a function of the divergence angle $\alpha$ for the $f$ = 0/$t_c$ case. The time-averaged flux $\left< \phi \right>$ is computed as the total volume of grains lost by the barchan during the considered interval, divided by the projected area $A$ and the time interval. We observe a lower flux in the cases when a flow change has taken place (i.e., when $\alpha$ $\neq$ 0$^\circ$), in agreement with the lower loss of grains shown in Fig. \ref{fig:grain_loss}a, but we cannot determine a clear trend for the variation of $\left< \phi \right>$ with $\alpha$. On the other hand, $\left< \phi \right>$ is considerably higher for $f$ = 10/$t_c$, as shown in Fig. \ref{fig:grain_motion}b, corroborating that the number of grains leaving the dune is higher for high frequencies ($1/f$ $<$ $t_c$). Figure \ref{fig:grain_motion}c shows the time-ensemble average of the longitudinal component of the grains' velocity $\left< v_x \right>$, normalized by $u_*$, as a function of $\alpha$, for $f$ = 0/$t_c$, from which we observe that velocities are roughly constant ($\left< v_x \right>$ $\approx$ 0.5$u_*$), with the exceptions of $\alpha$ = 10$^\circ$ and $\alpha$ = 180$^\circ$. For higher oscillation frequencies, values of $\left< v_x \right>$ (averages computed for each cycle) are slightly lower than in the continuous case, varying within 0.3$u_*$ and 0.45$u_*$, as shown in Figure \ref{fig:grain_motion}d as a function of time. PDFs of $V_x$, as well as dimensional forms of Figs. \ref{fig:grain_motion}c  and \ref{fig:grain_motion}d are available in the Supplementary Material.

In terms of resultant forces acting on grains, Fig. \ref{fig:grain_motion}e shows time-ensemble averages of the longitudinal component $\left< F_x \right>$ as a function of $\alpha$ for the continuous case, and we observe a non-monotonic behavior, with a minimum at $\alpha$ = 45$^\circ$ in which $\left< F_x \right>$ has approximately the same value of that for $\alpha$ = 0$^\circ$ ($\left< F_x \right>$ $\approx$ 1.2 $\times$ 10$^{-8}$ N). For high frequencies ($1/f$ $<$ $t_c$), values of $\left< F_x \right>$ (averages computed for each cycle) decrease with time, going from values of the other of those for low frequencies ($1/f$ $>$ $t_c$) to $\left< F_x \right>$ $\approx$ 0.5 $\times$ 10$^{-8}$ N, as shown in Fig. \ref{fig:grain_motion}f. This means that, although undergoing higher erosion rates, grains in the high-frequency case experience lower resultant forces.

\section{Conclusions}

\begin{sloppypar}
We carried out grain-scale numerical computations of a subaqueous barchan under bimodal flows, for which we varied both the angle $\alpha$ and frequency $f$ of oscillations. Different from previous experiments and simulations, we showed the typical trajectories, velocities, and resultant forces of moving grains. For $f$ $<$ $1/t_c$, where $t_c$ is the characteristic time for barchan formation, we found that the dune adapts to the new flow direction and recovers eventually the barchan shape, while for $f$ $>$ $1/t_c$ the dune has not enough time for adaptation and gives origin to more round barchans that are similar to the occluded dunes found on Mars. In addition, for $f$ $<$ $1/t_c$ the dune loses less grains than when under one-directional flows, while for $f$ $>$ $1/t_c$ (occluded-like dune) it loses more grains. Finally, we show that the resultant forces experienced by moving grains are lower for higher frequencies, with roughly constant average velocities and more asymmetric trajectory lines, the latter explaining the round shape reached after 4-5 cycles. For $f$ $<$ $1/t_c$, longitudinal forces vary non-monotonically with the angle, the grains experiencing lower resultant forces when $\alpha$ = 45$^\circ$. \corr{We note that our simulations were carried out for the subaqueous case, so that extrapolations to the terrestrial (eolian) and Martian cases must be done with caution. We also note that the use of different frequencies might result in dome-like dunes, as shown by \citet{Parteli5}.} Our findings provide new insights into the dynamics of dunes under seasonal flows, \corr{showing that the occluded dunes found on Mars can be the result of high-frequency variations in the flow direction, and revealing details of the motion of individual particles.}
\end{sloppypar}


\section*{Declaration of competing interest}
The authors declare that they have no conflict of interest.

\section*{Data Availability}
\begin{sloppypar}
	Data (numerical setup and outputs) supporting this work were generated by ourselves and are available in Mendeley Data \cite{Supplemental2} under the CC-BY-4.0 license. The numerical scripts used to process the numerical outputs are also available in Mendeley Data \cite{Supplemental2} under the CC-BY-4.0 license.
\end{sloppypar}

\section*{Acknowledgments}
\begin{sloppypar}
	The authors are grateful to FAPESP (Grant Nos. 2018/14981-7, 2019/10239-7 and 2022/01758-3) and to CNPq (Grant No. 405512/2022-8) for the financial support provided. The authors thank Nicolao Cerqueira Lima for the help with the numerical setup.
\end{sloppypar}







\end{document}